\documentclass[%
aip,
amsmath,amssymb,
reprint,%
]{revtex4-1}

\usepackage{graphicx}% Include figure files
\usepackage{dcolumn}% Align table columns on decimal point
\usepackage{bm}% bold math
\usepackage[utf8]{inputenc}
\usepackage[T1]{fontenc}
\usepackage{mathptmx}
\usepackage{textcomp}
\usepackage{gensymb}
\usepackage{xcolor}

\begin{document}
	
\preprint{AIP/123-QED}
	
\title{Understanding Thermal Annealing of Artificial Spin Ice}
	% Force line breaks with \\
	
\author{Xiaoyu Zhang}
\affiliation{Department of Applied Physics, Yale University, New Haven, CT 06511, USA}
\affiliation{Department of Physics, University of Illinois at Urbana-Champaign, Urbana, IL 61801, USA}
\affiliation{Frederick Seitz Materials Research Laboratory, University of Illinois at Urbana-Champaign, Urbana, IL 61801, USA}

\author{Yuyang Lao}%
\affiliation{Department of Physics, University of Illinois at Urbana-Champaign, Urbana, IL 61801, USA}
\affiliation{Frederick Seitz Materials Research Laboratory, University of Illinois at Urbana-Champaign, Urbana, IL 61801, USA}

\author{Joseph Sklenar}
\affiliation{Department of Physics, University of Illinois at Urbana-Champaign, Urbana, IL 61801, USA}
\affiliation{Frederick Seitz Materials Research Laboratory, University of Illinois at Urbana-Champaign, Urbana, IL 61801, USA}
\affiliation{Department of Physics and Astronomy, Wayne State University, Detroit, MI 48201, USA }

\author{Nicholas S. Bingham}
\affiliation{Department of Applied Physics, Yale University, New Haven, CT 06511, USA}

\author{Joseph T. Batley}
\affiliation{Department of Chemical Engineering and Materials Science, University of Minnesota, Minneapolis, MN 55455, USA}

\author{Justin D. Watts}
\affiliation{Department of Chemical Engineering and Materials Science, University of Minnesota, Minneapolis, MN 55455, USA}
\affiliation{School of Physics and Astronomy, University of Minnesota, Minneapolis, MN 55455, USA}

\author{Cristiano Nisoli}
\affiliation{Theoretical Division and Center for Nonlinear Studies, MS B258, Los Alamos National Laboratory, Los Alamos, NM 87545, USA}

\author{Chris Leighton}
\affiliation{Department of Chemical Engineering and Materials Science, University of Minnesota, Minneapolis, MN 55455, USA}

\author{Peter Schiffer}
\email{peter.schiffer@yale.edu}
\affiliation{Department of Applied Physics, Yale University, New Haven, CT 06511, USA}
\affiliation{Department of Physics, University of Illinois at Urbana-Champaign, Urbana, IL 61801, USA}
\affiliation{Frederick Seitz Materials Research Laboratory, University of Illinois at Urbana-Champaign, Urbana, IL 61801, USA}
\affiliation{Department of Physics, Yale University, New Haven, CT 06511, USA}

\date{\today}% It is always \today, today,

\begin{abstract}
	We have performed a detailed study of thermal annealing of the moment configuration in artificial spin ice.  Permalloy (Ni$_{80}$Fe$_{20}$) artificial spin ice samples were examined in the prototypical square ice geometry, studying annealing as a function of island thickness, island shape, and annealing temperature and duration.  We also measured the Curie temperature as a function of film thickness, finding that thickness has a strong effect on the Curie temperature in regimes of relevance to many studies of the dynamics of artificial spin ice systems.  Increasing the interaction energy between island moments and reducing the energy barrier to flipping the island moments allows the system to more closely approach the collective low energy state of the moments upon annealing, suggesting new channels for understanding the thermalization processes in these important model systems. 
\end{abstract}

\maketitle

Artificial spin ice systems are two-dimensional arrays of nanoscale elements, typically composed of single domain ferromagnetic islands\cite{wang06}. These systems have been the subject of extensive study and have provided models for the study of a range of novel collective behaviors\cite{gilbert16}. Certain artificial spin ice geometries have well-defined collective magnetic ground states, such as the square lattice\cite{wang06}, while others have intrinsically disordered and complex ground states, such as the Shakti lattice\cite{morrison13,gilbert14,lao18}.  These low-energy collective states have sparked considerable interest in attempting to realize the lowest energy state of different artificial spin ice lattices\cite{ke08,morgan11,farhan13,drisko15,morley18}. One successful approach to collective energy minimization involves annealing the arrays by heating them to temperatures near or above the Curie temperature ($T_C$) of the ferromagnetic material\cite{porro13,zhang13}.  Upon cooling, the island moments arrange themselves into a low energy state via magnetostatic interactions. Using this method, both long-range-ordered\cite{zhang13,porro13,sklenar}  and intrinsically disordered ground states\cite{perrin16,gilbert14}  have been achieved, both in permalloy (Ni$_{80}$Fe$_{20}$) and in other alloys \cite{drisko15,morley18}. Notably, the method works well even for geometries known to exhibit slow relaxation toward the low energy state\cite{gilbert14}.  Given the high $T_C$ of permalloy, and its importance as a model material for these systems, we investigated thermal annealing of permalloy artificial spin ice by varying the annealing conditions and the geometry of the islands, with the goal of understanding how to improve the effectiveness of annealing. 

We fabricated our artificial square spin ice samples on Si wafers coated with a 200-nm-thick layer of Si-N deposited by low pressure chemical vapor deposition.  The nanoislands, with varied lateral dimensions and inter-island gaps indicated below, were produced by electron beam lithography and lift-off as described previously\cite{zhang13}. The total area of all nanoislands in each square ice sample was about 200$\times$200 $\mu m^2$. In order to keep uniformity of all nanoislands, the write field for lithography was set to cover the whole sample. We deposited our samples with various thicknesses of permalloy (2 – 100 nm) and a 3 nm Al cap layer to prevent oxidation in ultrahigh vacuum by molecular beam deposition (using electron beam evaporation). The thickness was established via a quartz-crystal-based rate monitor, which was carefully calibrated by performing grazing-incidence X-ray reflectivity measurements. We performed thermal annealing using a programmable heater (HeatWave Lab) in a vacuum chamber with a base pressure of approximately ${10}^{-6}$ Torr in order to prevent sample degradation via oxidation, lateral diffusion, and delamination.  After each annealing cycle, we characterized the magnetic state of the arrays in two locations using a magnetic force microscope (MFM), yielding a corresponding microstate map of the arrays.  For measurement of the $T_C$ of the permalloy, we measured the temperature dependence of the magnetization of continuous permalloy films in a SQUID vibrating sample magnetometer (MPMS3 from Quantum Design); the base pressure during $T_C$ measurements was approximately ${10}^{-3}$ Torr. The permalloy films used for $T_C$ measurements had an extra 100 nm Si-N encapsulation layer on top deposited by low pressure chemical vapor deposition to protect them from degradation during the measurements.

\begin{figure}
	\includegraphics[width=\linewidth]{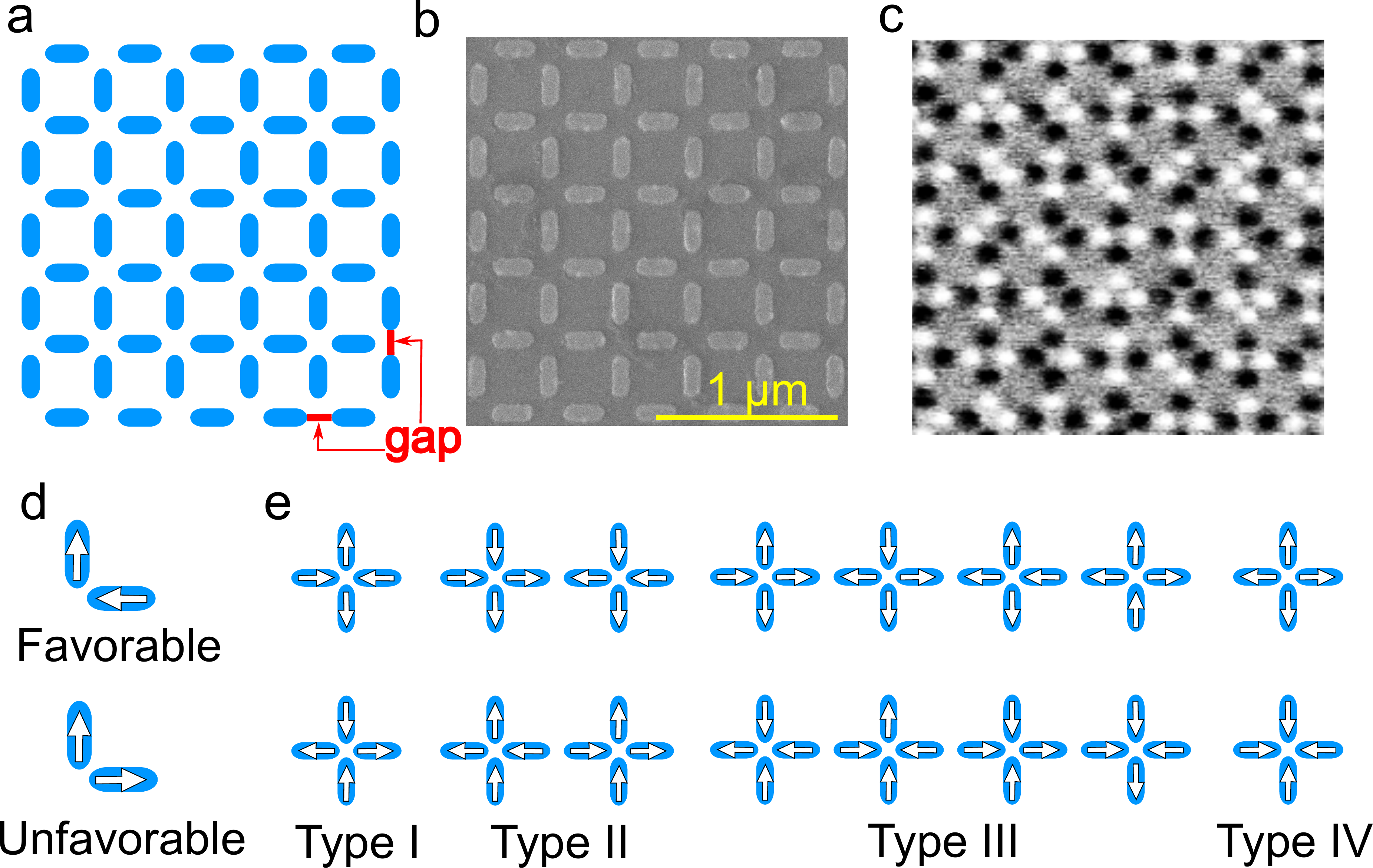}% Here is how to import EPS art
	\caption{\label{schematic} (a) Schematic of the square artificial spin ice lattice geometry.  The gap is defined as the edge-to-edge distance between a pair of next nearest neighbor islands. (b) SEM image and (c) MFM image of the square artificial spin ice lattice. (d) The favorable and unfavorable alignment for moments of the nearest neighbor islands. (e) The energy hierarchy of the vertex types in square artificial spin ice with increasing vertex energy from Type I to Type IV.}
\end{figure}

The artificial square ice lattice is shown in Figure \ref{schematic} with typical SEM (scanning electron microscope) and MFM images shown in Figure \ref{schematic}b and \ref{schematic}c.  Artificial spin ice lattices are often analyzed in terms of the vertices of the lattices, defined as the junctions of multiple islands; the collective magnetic energy is determined to a good approximation by the sum of the magnetostatic energy associated with interactions among the moments at the vertices.  The interaction energy between islands can be modified by varying the island shape and the gap length of the lattice (defined as the edge-to-edge distance between two next nearest neighbor islands (Figure \ref{schematic}a)). 

Each square artificial spin ice vertex consists of four islands with 16 possible magnetic moment configurations (Figure \ref{schematic}e) that can be classified into four types.Unlike the three-dimensional pyrochlore spin ice structure\cite{castelnovo12}, in which all four spins are equidistant resulting in six degenerate two-in-two-out ground states for each tetrahedron, the two-dimensional square artificial spin ice has four islands that have stronger interactions between the nearest neighbor pairs and weaker interactions between the next nearest neighbor pairs. Therefore, the six-fold ground state degeneracy is broken in square artificial spin ice, and the two Type I vertex states are the only ground states for individual vertices. The collective ground state of the system then contains only Type I vertices in a long-range-ordered antiferromagnetic configuration\cite{moller06}.  This makes the system ideal for studying the effects of thermal annealing, since the ground state fraction provides a clear measure of the effectiveness of annealing.

All annealing treatments on the square ice samples were performed after magnetizing the samples along a 45$^{\circ}$ angle relative to the lattice to ensure that all the vertices were initially set to a Type II state.  We then heated the samples to the annealing temperature at a rate of 10 K per minute and held the samples at the annealing temperature for 10 minutes (unless otherwise stated).  Finally, they were cooled at 1 K per minute back to 673 K (low enough that the moments were thermally stable), then quickly cooled to room temperature in about 90 minutes.  For each annealing protocol, we repeated the measurement up to four times.  We then measured the effectiveness of annealing for various parameter values, as quantified through the Type I vertex population fraction, i.e., the ground state fraction, after annealing\cite{drisko15}. The error bars below are taken as the standard deviation of vertex fractions from all MFM images.  The total number of MFM images for each corresponding data point is two times the number of repeated experiments, as we took two MFM images after each annealing cycle.  Acquired MFM images typically spanned 1700 islands over an area of 15$\times$15 $\mu$m$^2$. 

Since thermal annealing of artificial spin ice moments requires approaching the ferromagnetic $T_C$, we measured the ferromagnetic moment as a function of temperature in continuous films to determine $T_C$.  In thin film ferromagnetic systems, the second-order transition to the ordered state shows a strong dependence on film thickness associated with finite size scaling\cite{allan70,fisher72},  and we thus measured a range of film thicknesses. The magnetization of the films was measured, both zero-field-cooling (ZFC) and field-cooling (FC), as a function of temperature in an in-plane magnetic field of 100 Oe.  We estimated T$_C$ as the point of inflection in the magnetization versus temperature curve (raw data used to determine $T_C$ are provided in Supplementary Information, Figure SI-1). Although degradation effects such as delamination did not impact our annealing of artificial spin ice samples in high vacuum, we did observe degradation of artificial spin ice samples annealed in the presence of low gas pressures, as in our magnetometer\cite{laothesis}. Notice that we added a Si$_3$N$_4$ encapsulation layer on the continuous permalloy films samples used for the $T_C$ measurements to prevent such degradation. 

\begin{figure}
	\includegraphics[width=0.8\linewidth]{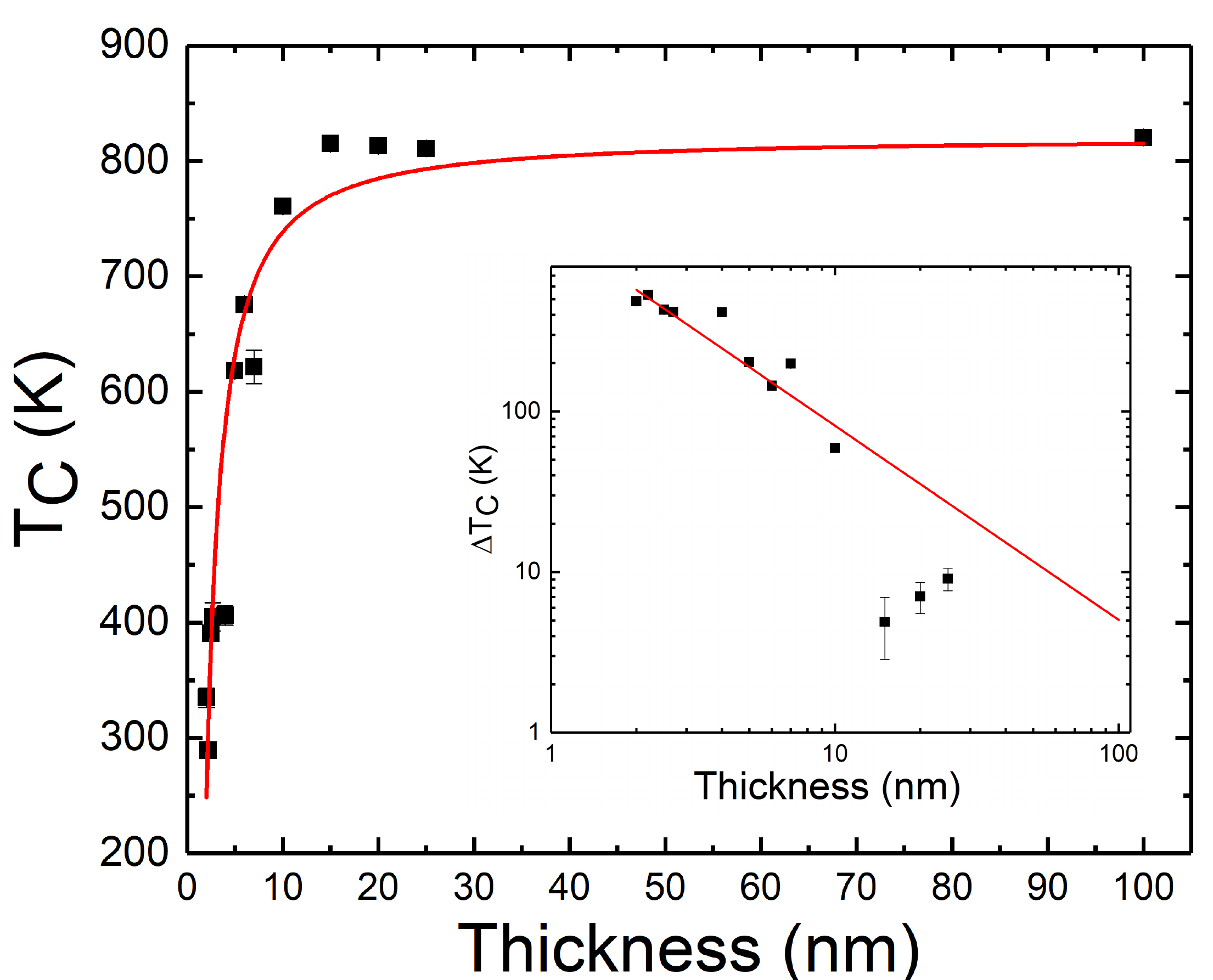}% Here is how to import EPS art
	\caption{\label{Tc} Thickness dependence of the Curie temperature for permalloy films.  The error bars represent the mean absolute error of $T_C$ extracted from ZFC and FC measurements. The red curve shows the fit for equation \ref{finitesize}. The inset shows the same data, plotted as $\Delta T_C$, which is defined as the difference between the bulk $T_C$ and the measured thin film $T_C$, assuming a bulk $T_C$ of 820 K}.
\end{figure}

The data in Figure \ref{Tc} reveal a strong thickness dependence to $T_C$, particularly below 10 nm.  The inset to Figure \ref{Tc} plots the reduction in $T_C$ from bulk ($T_C$) vs. the thickness.  The red line is a fit to the function :

\begin{eqnarray}
\frac{T_C (bulk)-T_C (t)}{T_C (bulk)}=\big(\frac{\xi_0}{t}\big)^{\lambda} \label{finitesize}
\end{eqnarray}

where t is the thickness of the permalloy film, $T_C(t)$ is the Curie temperature at thickness t, and $T_C  (bulk)$ is the Curie temperature of bulk permalloy (taken here to be 820 K\cite{Bozorth93}).  The fit gives a correlation length $\xi_0$ = 1.48 nm $\pm$ 0.11 nm and a critical shift exponent $\lambda$ = 1.21 $\pm$ 0.12, in reasonable agreement with previous measurements of ultra-thin permalloy films ($\lambda$ = 1.04) using spin-polarized cascade electrons\cite{marui89}. 

This thickness dependence of the permalloy $T_C$ has clear implications for artificial spin ice studies.  For example, the suppressed $T_C$ for thinner films is directly relevant to the many photoelectron emission microscopy (PEEM) measurements that have been conducted on permalloy artificial spin ice\cite{lao18,farhan13,farhan14,gliga17}.  Those measurements focused on films of thickness around 3 nm, and they are therefore conducted relatively close to $T_C$ of the substituent ferromagnetic material.  Our data suggest that the observed dynamics in the moment fluctuations might be significantly influenced by the reduced moment associated with the proximity to $T_C$.

We now discuss studies of annealing while varying different parameters of the lattices and the annealing process.  We first describe results of annealing with varying maximum annealing temperature.  As the temperature approaches the ferromagnetic transition, the magnetization of permalloy islands decreases, resulting in a decrease of the energy barrier to changing the magnetization orientation of each island along with increased thermal energy in the system.  When thermal energy is large enough to overcome the energy barrier required for a moment to switch, the island moments become thermally active and begin to organize themselves into an energetically preferred configuration.  The ground-state vertices presumably nucleate in islands with slightly lower energy barriers (potentially associated with imperfections in the lithography), and then expand to large domains.  This allows annealing to occur even below the $T_C$ of continuous permalloy films.  

\begin{figure}
	\includegraphics[width=\linewidth]{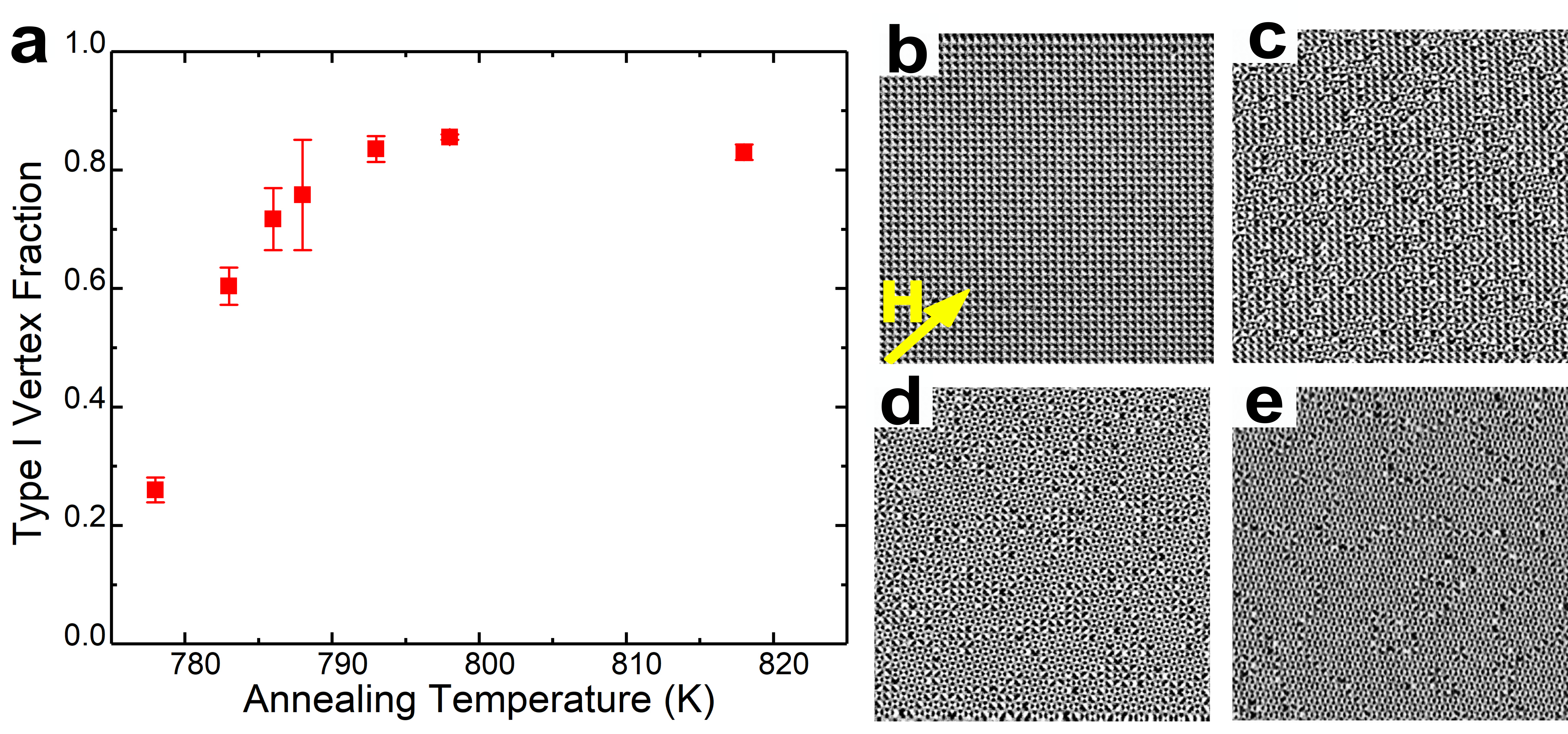}% Here is how to import EPS art
	\caption{\label{maxT} (a) The ground state fraction as a function of the maximum temperature of annealing for lattices with 160$\times$60 nm$^2$ island shape, 180 nm gap, and 25 nm thickness. (b)-(e) 15$\times$15 $\mu m^2$ MFM images: (b) shows a completely polarized state after a 1000 Oe field is applied along the arrow direction, (c) shows 25$\%$ Type I vertices after annealing at 778 K, (d) shows 70$\%$ Type I vertices after annealing at 786 K, and (e) shows 83$\%$ Type I vertices after annealing at 818 K.}
\end{figure}

Our annealing-temperature-dependent data, taken on samples with island shape of 160$\times$60 nm$^2$, gap of 180 nm, and thickness of 25 nm, are shown in Figure \ref{maxT}a, which plots the Type-I vertex fraction.  As expected, we observed a rather sharp increase in ground state fraction as the temperature is increased towards $T_C$, with the Type I vertex population reaching a maximum at about 83$\%$ (see Figure \ref{maxT}a and Figure \ref{maxT}e) at approximately 790 K.  This increase in Type-I vertex fraction corresponds to a decrease in the residual polarized vertices as the annealing temperature was increased (see Supplemental Information, Figure SI-2). The combination of these data indicates that the moments in the array become thermally active over a range of temperatures.  This has the implication that annealing can only succeed in allowing the full system to explore the possible variations in moment configurations by exceeding a certain temperature threshold, where all moments are thermally active. For all data discussed below, there was no significant residual population of the initially polarized vertices, indicating that we exceeded that threshold temperature, and all the moments had the opportunity to thermally fluctuate during the annealing process.

We also examined the dwell time dependence by holding a sample for 10 min, 60 min, or 120 min at an annealing temperature of 773 K or 783 K (data shown in Supplementary Information, Figure SI-3).  We did not observe any significant effect on the vertex population density, most likely indicating that the system reaches thermal equilibrium on a time scale faster than our experimental temporal resolution, although we cannot rule out the possibility of very slow dynamics associated with long relaxation time scales near the superparamagnetic blocking temperature.  

\begin{figure}
	\includegraphics[width=\linewidth]{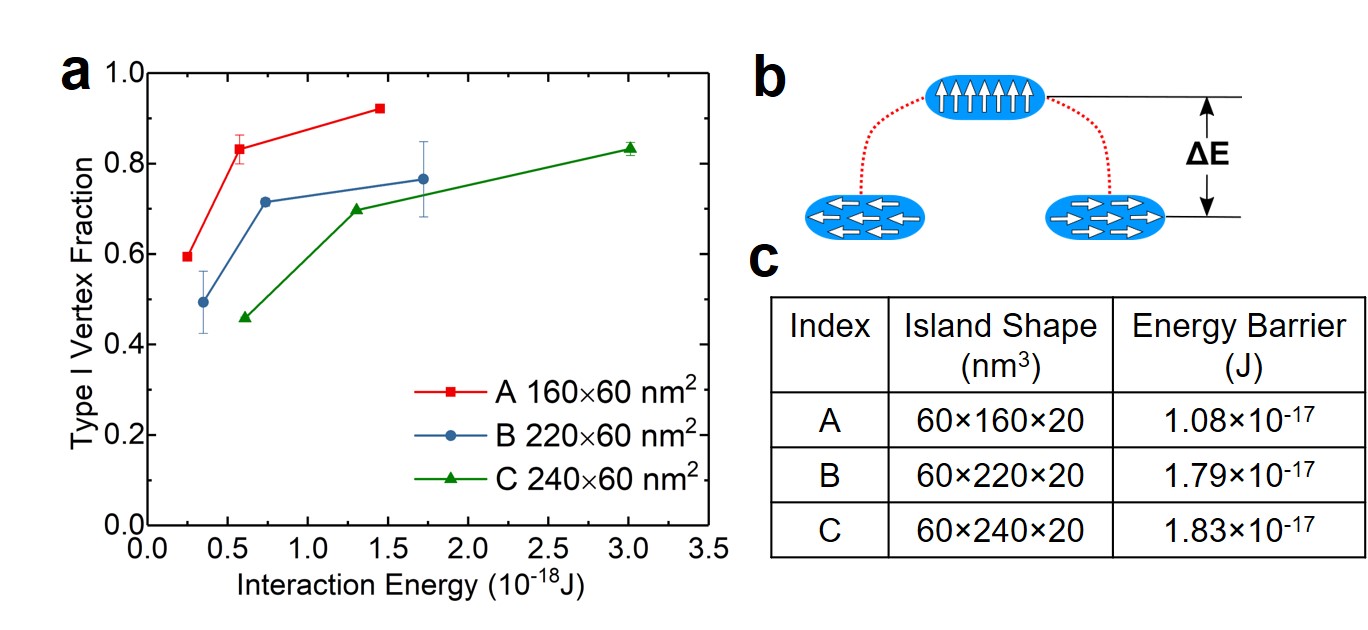}% Here is how to import EPS art
	\caption{\label{shape} (a) Island shape-dependence of the ground state fraction on 20-nm-thick lattices with various island shapes and gaps after annealing at 783 K for 10 minutes. (b) Schematic illustrating how the energy barrier is approximated. (c) Approximate energy barriers for 20 nm-thick and 60 nm-wide islands of various lengths.}
\end{figure}

We next investigated the island shape dependence of the ground state fraction after annealing by using a 20-nm-thick sample with island dimensions 160$\times$60 nm$^2$, 220$\times$60 nm$^2$, and 240$\times$60 nm$^2$, with gaps of 100 nm, 180 nm, and 280 nm, as shown in Figure \ref{shape}.  Because these arrays varied in both shape and spacing, we plot the results against the interaction energy between nearest neighbor island pairs, calculated with the micromagnetics code OOMMF\cite{oommf}  at zero temperature.  The samples were annealed at a temperature of 783 K, i.e., below $T_C$ at this thickness.  This choice of maximum temperature allowed us to more precisely control the annealing process than would have been possible by annealing above $T_C$ and cooling at rates that necessarily depend on the thermal coupling of the sample to the heater stage.  Figure \ref{shape}a presents a plot of Type I vertex population fraction versus interaction energy for various island shapes.  As previously observed\cite{zhang13}, the ground state fraction is increased with greater interaction strength, a natural consequence of the energy scale of interactions relative to the superparamagnetic blocking temperature. Furthermore, structural disorder intrinsic to lithography will lead to a distribution of both blocking temperature and local interaction strength, which will disrupt the system’s ability to reach a collective low-energy state; stronger coupling should overcome this effect. 

The annealing process was also more effective with smaller island aspect ratio.  Since these permalloy films have in-plane magnetization and weak magnetocrystalline anisotropy, the aspect ratio dictates the energy barrier to flipping each island moment, via shape anisotropy.  The energy barrier can be roughly parameterized as the energy difference between configurations with moments along the long axis and the short axis (Figure \ref{shape}b).  This energy barrier for each island shape was calculated using MUMAX3\cite{mumax3}  and is shown in Figure \ref{shape}c (note that these values are calculated assuming the magnetization at zero temperature).  We see that the increase in ground state fraction indeed corresponds to decreases in this energy barrier. We obtained qualitatively consistent results for two series of islands with 80 nm width and varying length, suggesting that the lower barrier is associated with more complete annealing of the moments into a low energy state.

\begin{figure}
	\includegraphics[width=\linewidth]{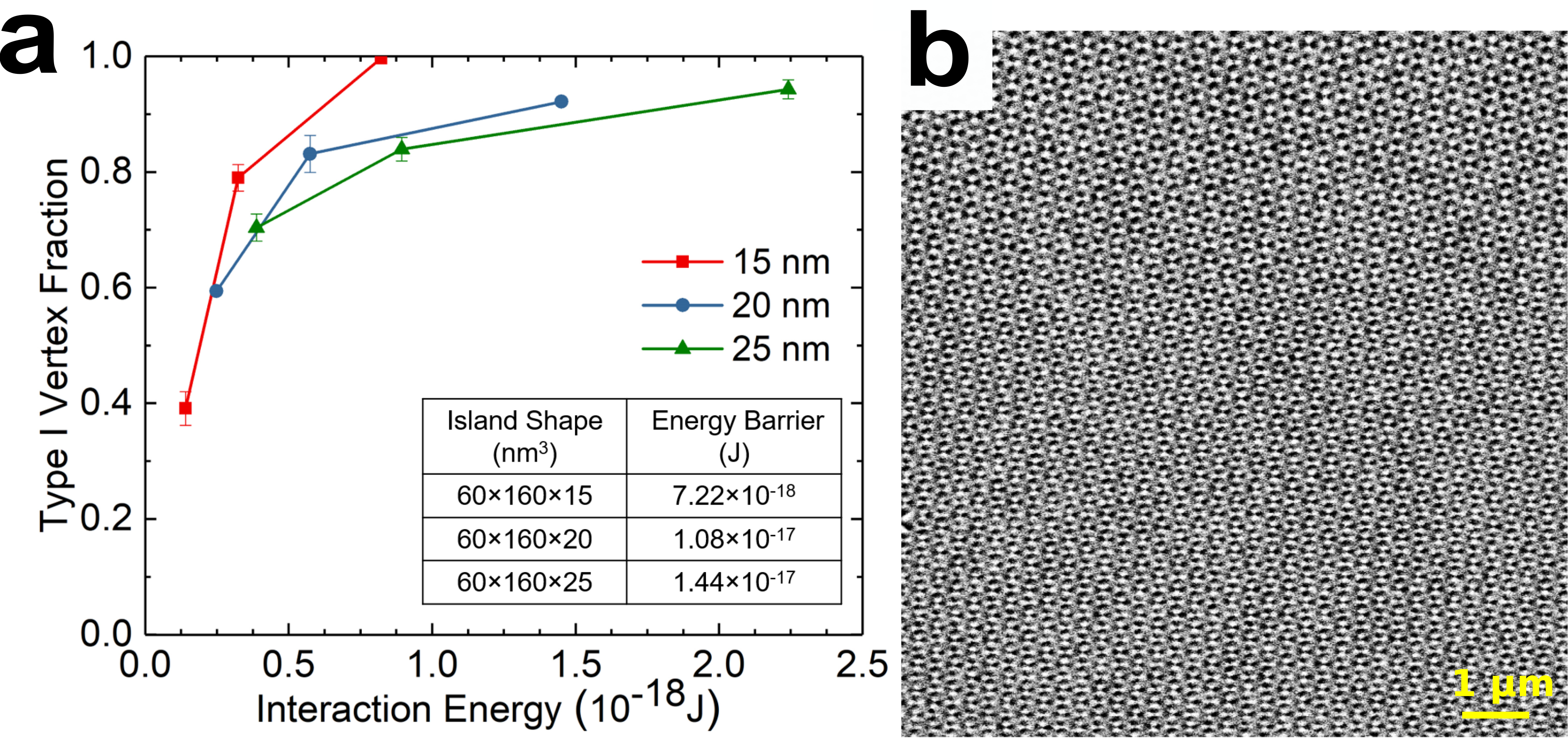}% Here is how to import EPS art
	\caption{\label{thick} Thickness-dependent ground state fraction for 160$\times$60 nm$^2$ square lattices with gaps of 100 nm, 180 nm, and 280 nm after annealing at 783 K for 10 minutes. The inserted table shows the energy barrier for each island thickness.  (b) 10$\times$10 $\mu m^2$ MFM image of 160$\times$60 nm$^2$ island shapes with 100 nm gap and 15 nm thickness.  A near-ideal ground state is achieved.}
\end{figure}

We next examined the ground state fraction for annealed square artificial spin ice arrays with various thicknesses (15 nm, 20 nm, and 25 nm).  We chose these thicknesses to be large enough that the island moments were unaffected by the MFM tip, allowing the moment configuration to be studied after annealing.  These are also thicknesses for which the $T_C$ appears approximately constant (see Figure \ref{Tc}).  For these measurements, we studied island dimensions of 160$\times$60 nm$^2$ with gaps of 100 nm, 180 nm, and 280 nm.  The post-annealing ground state fraction is plotted versus interaction energy in Figure \ref{thick}a, and the energy barriers for various thickness islands are shown in the inset.  These data again suggest that the ground state fraction for a given interaction energy strength is correlated with decreasing the size of the approximate energy barrier to island moment reversal.  In fact, the highest ground state fraction in this study is seen for the smallest aspect ratio islands with 160$\times$60 nm$^2$ lateral dimensions, 100 nm gap, and 15 nm thickness, for which an MFM image is shown in Figure \ref{thick}b. 

Artificial spin ice systems near their $T_C$ are in a fascinating regime where the magnetization is strongly temperature dependent, with corresponding impacts on both the interisland interactions and the superparamagnetic moment relaxation time.  As a result, the collective thermal dynamics of the moments in this regime, and how that relates to the intrinsic frustration for which these systems are designed, cannot be analyzed within a framework of simple superparamagnetic relaxation.  Our data on the fraction of ground state vertices serves as a proxy for how well annealing accesses the complex many-body states of artificial spin ice systems.  The results indicate that accessing the lowest energy collective states by annealing can be improved by appropriate choice of island shape, spacing, thickness, and annealing temperature, but that some systems are not amenable to reaching the lowest possible collective energy state by annealing methods.  Previous annealing studies also failed to attain a complete ground state for certain lattices\cite{zhang13,drisko15}, and the ordered ground state of the kagome lattice has never been achieved by annealing\cite{anghinolfi15}.  These results suggest that the effectiveness of annealing is limited by local minima in the energy landscape associated with lithographic disorder\cite{drisko15}, and that lithographic perfection might be a limiting factor for this approach.  On the other hand, given the complexity of the many-body relaxation process in a system where temperature dependence to relaxation times is associated with multiple factors, we cannot rule out limits associated with the collective relaxation process or very long relaxation time scales that are outside the scope of the present work.  

Finally, we note that our results on the strong $T_C$ dependence on the film thickness has implications for other studies of permalloy artificial spin ice, and PEEM measurements in particular, since those measurements appear to be in a regime in which the magnetization has considerable temperature dependence.  This regime should thus provide fertile ground for interesting new physical phenomena in nanostructured systems.  While we studied only the square ice lattice, we expect these considerations to carry forward to other geometries, opening the possibility of more closely examining low energy states and collective thermal relaxation processes in a range of systems.

\begin{acknowledgments}
This work at the University of Illinois at Urbana-Champaign and Yale University was funded by the US Department of Energy, Office of Basic Energy Sciences, Materials Sciences and Engineering Division under Grant No. DE-SC0010778. This work was carried out in part in the Frederick Seitz Materials Research Laboratory Central Research Facilities, University of Illinois at Urbana-Champaign.  Work at the University of Minnesota was supported by NSF through DMR-1807124.  The work of CN was carried out under the auspices of the US DoE through LANL, operated by Triad National Security, LLC (Contract No. 892333218NCA000001) and financed by DoE LDRD.
\end{acknowledgments}

\section*{Supplementary Information}
\renewcommand{\thefigure}{SI-\arabic{figure}}
\setcounter{figure}{0}

\begin{figure*}
	\includegraphics[width=\linewidth]{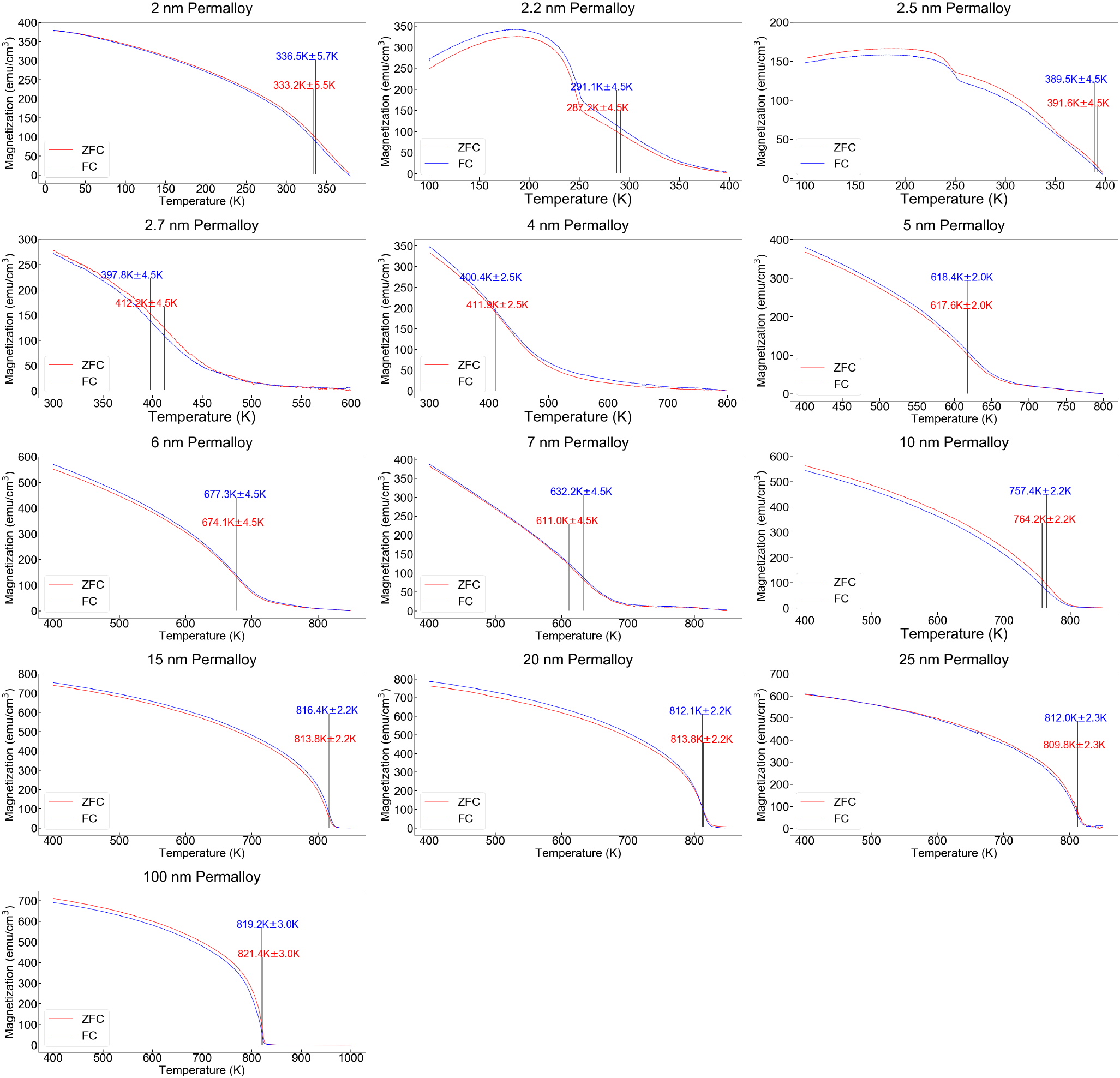}% Here is how to import EPS art
	\caption{\label{mt}Magnetization versus temperature curves, labeled with the deduced $T_C$, for illustrative 2 nm, 2.2 nm, 2.5 nm, 2.7 nm, 4 nm, 5 nm, 6 nm, 7 nm, 10 nm, 15 nm, 20 nm, 25 nm, and 100 nm thickness permalloy films.}
\end{figure*}

\textbf{Magnetization data:} We measured the magnetization of continuous permalloy films of different thickness, both zero-field-cooled and field-cooled, as a function of temperature (Figure \ref{mt}).  Each permalloy film had a 100 nm Si-N encapsulation layer deposited by low pressure chemical vapor deposition to prevent degradation during high temperature measurement.  A 100 Oe in-plane field was applied for each measurement. The error on $T_C$ represents the uncertainty in finding the inflection point in magnetization vs. temperature.  All measurements were performed in a SQUID vibrating sample magnetometer.  Note that some of the features near 250 K in the thinnest films may be associated with a parasitic oxide phase.

\begin{figure}
	\includegraphics[width=\linewidth]{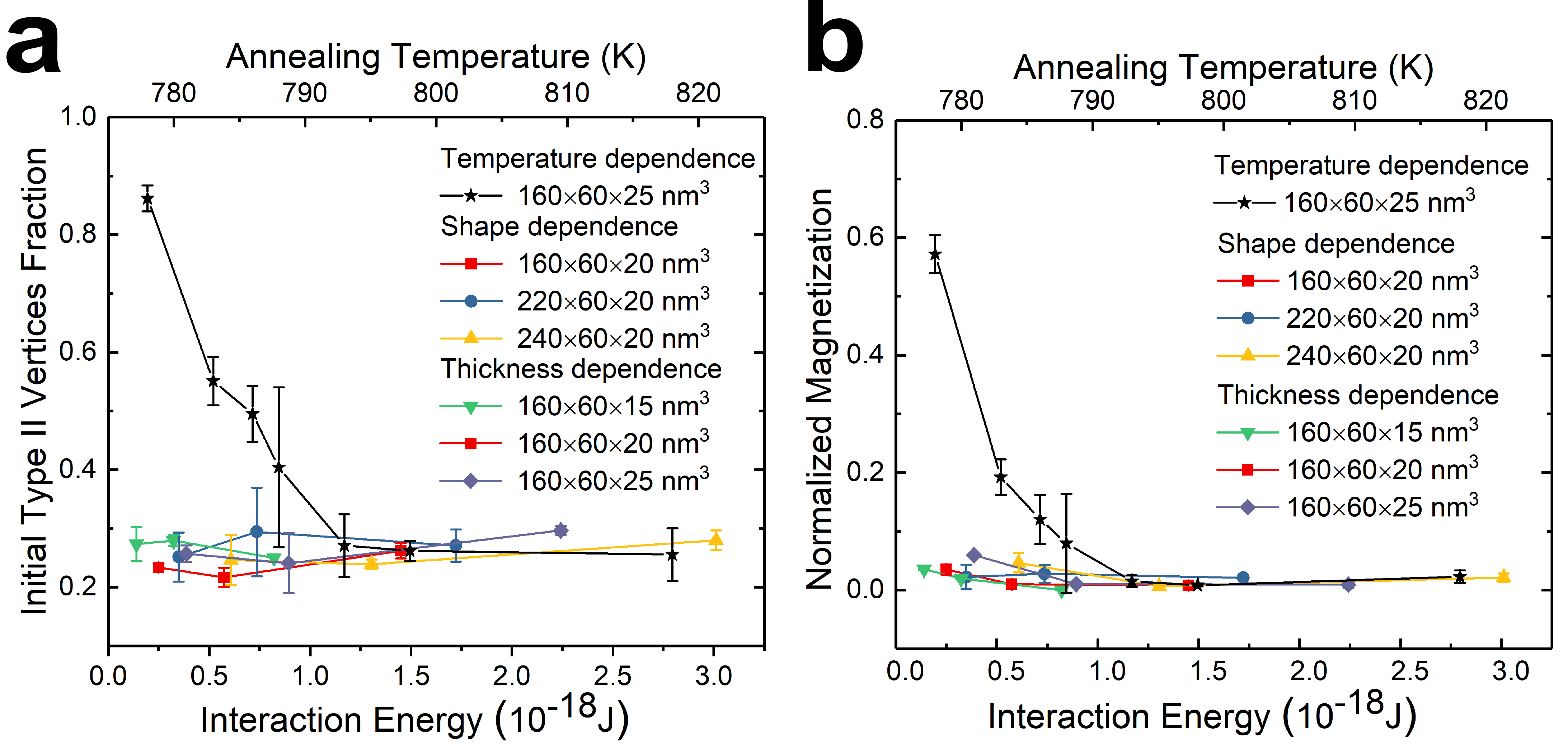}% Here is how to import EPS art
	\caption{\label{si} The fraction of unchanged type II vertices (a) and normalized magnetization (b) as a function of interaction energy for shape-dependent and thickness-dependent studies, or as a function of temperature for the annealing temperature study.}
\end{figure}

\textbf{Residual polarization data:} To test whether square ice lattices were truly thermalized by our annealing treatments, i.e., whether all the island moments had the opportunity to reverse due to thermal fluctuations, we measured the initial Type II vertex fraction (Figure \ref{si}a) and normalized	magnetization (Figure \ref{si}b) for all MFM images used for shape-dependence, thickness dependence, and annealing temperature dependence studies. As all square ice lattices were polarized with Type II states initially, we counted the number of unchanged vertices, i.e., vertices that had the initial polarization, and compared that with the total number of Type II vertices. We calculated the fraction of initial Type II vertices, defined as the ratio of unchanged vertices to all Type II vertices. The fraction of Type II vertices that retain their initial polarization is close to 25$\%$ for all annealing data, except for the temperature-dependence data associated with the measurement of the annealing temperature below 790 K (Figure 3in the main paper). This indicates that possibilities for each kind of type II configuration are equal, thus the vertices didn't have memory of their initial states after thermalization, except for the data in Figure 3.

The net magnetization from the MFM images was calculated by summing the moments from individual islands. We normalized the net magnetization to the saturation magnetization of the completely polarized state. Again the magnetization is close to zero for all annealing data, except for those data associated with the measurement of the annealing temperature below 790 K. This shows that the islands were able to reverse magnetization during the annealing processes for all measurements except those low-temperature annealing data shown in Figure 3.

\begin{figure}
	\includegraphics[width=0.8\linewidth]{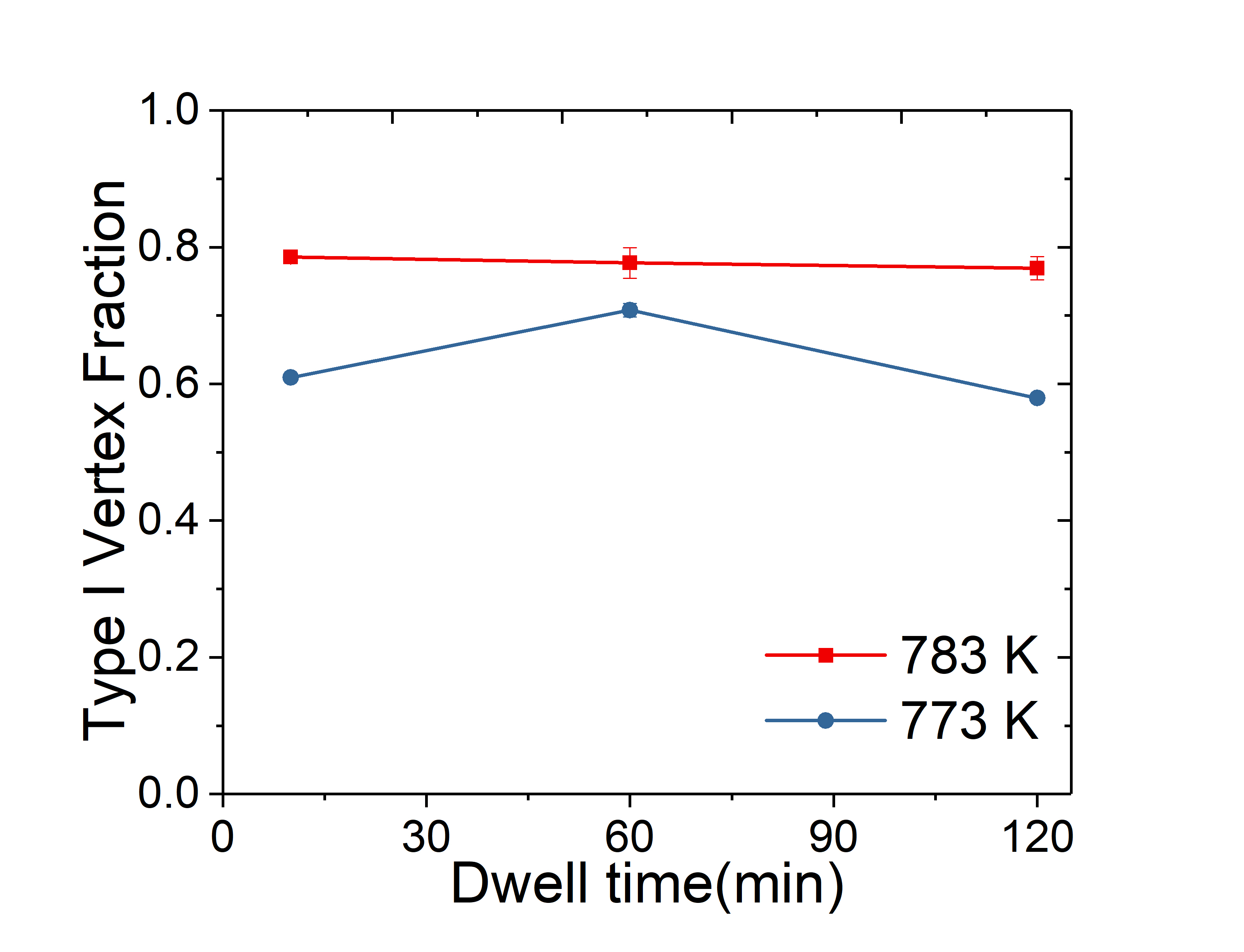}% Here is how to import EPS art
	\caption{\label{dwell} (a) The type I vertex fraction as a function of dwell time for annealing of 160$\times$60 nm$^2$ island shapes with 180 nm gap and 25 nm thickness, at 773 K and 783 K.}
\end{figure}

\textbf{Dwell time dependence:} We examined dwell time dependence in annealing by holding the sample at the maximum temperature during annealing for different periods of time.  Our data were taken on square ice arrays with island dimensions of 160$\times$60 nm$^2$, a gap of 180 nm, and thickness of 25 nm.  The sample was held at 773 K or 785 K for 10 min, 60 min, and 120 min.  As shown in Figure \ref{dwell}, there is no significant change in Type I vertex population fraction for different dwell times. 

\nocite{*}
\bibliography{citations}

\end{document}